# A comparison of emulation methods for Approximate Bayesian Computation

Franck Jabot[a], Guillaume Lagarrigues[b], Benoît Courbaud[b], Nicolas Dumoulin[a]

[a]Irstea, UR LISC, Laboratoire d'Ingénierie pour les Systèmes Complexes, 9 avenue Blaise Pascal, CS 20085, 63178 Aubière, France.

[b]Irstea, UR EMGR, Ecosystèmes Montagnards, 2 rue de la Papeterie, BP76, 38402 Saint-Martin-d'Hères Cedex, France.

# Abstract

Approximate Bayesian Computation (ABC) is a family of statistical inference techniques, which is increasingly used in biology and other scientific fields. Its main benefit is to be applicable to models for which the computation of the model likelihood is intractable. The basic idea of ABC is to empirically approximate the model likelihood by using intensive realizations of model runs. Due to computing time limitations, ABC has thus been mainly applied to models that are relatively quick to simulate. We here aim at briefly introducing the field of statistical emulation of computer code outputs and to demonstrate its potential for ABC applications. Emulation consists in replacing the costly to simulate model by another (quick to simulate) statistical model called emulator or meta-model. This emulator is fitted to a small number of outputs of the original model, and is subsequently used as a surrogate during the inference procedure. In this contribution, we first detail the principles of model emulation, with a special reference to the ABC context in which the description of the stochasticity of model realizations is as important as the description of the trends linking model parameters and outputs. We then compare several emulation strategies in an ABC context, using as case study a stochastic ecological model of community dynamics. We finally describe a novel emulation-based sequential ABC algorithm which is shown to decrease computing time by a factor of two on the studied example, compared to previous sequential ABC algorithms. Routines to perform emulation-based ABC were made available within the R package EasyABC.

# 1. Introduction

A number of ecological and environmental models represent detailed mechanistic or interaction processes to grasp the complexity of living biological systems (Grimm and Railsback 2005). These models pose important methodological challenges to be calibrated with field data, since available data often inform only indirectly on the processes themselves. For instance, population-level information may be available rather than individual-level data, or moving individuals may not be tracked through space and time. In such cases, the use of inverse methods is necessary to infer from indirect data the processes that have most influenced the system studied. Historically, pattern-oriented modelling has played a central role for diffusing this concept of inverse modelling in ecology (Grimm et al. 2005). This field of research has been linked to the more general Bayesian framework in the last decade through the development of a family of computational inference techniques called Approximate Bayesian Computation (ABC, Beaumont et al. 2002) or likelihood-free inference. These techniques are particularly well suited to perform model inference when the likelihood of the model is analytically intractable, since ABC empirically estimates this likelihood by relying on intensive model simulations.

ABC consists in simulating a large number of times the model with parameters $\theta$ drawn in a prior distribution $\pi(\theta)$. The results of each simulation $x \sim f(x|\theta)$ are then generally summarized by a number of summary statistics $S(x)$, although this step has been bypassed in some ABC applications (Sousa et al. 2009). The same summary statistics are computed on the data y to be fitted. A distance between data and simulations is computed $\rho(S(x),S(y))$, and the simulations at a distance smaller than a threshold level $\varepsilon$, called tolerance, are retained. The parameters values used for these retained simulations form a sample of the approximate posterior distribution, whose density is proportional to $\pi(\theta)\, pr_\theta\{\rho(S(x),S(y))<\varepsilon\}$ where $pr_\theta\{z\}$ represents the probability distribution of z for a given parameter $\theta$. This approximate posterior distribution approximates the exact posterior distribution which is proportional to $\pi(\theta) \, pr_\theta\{x=y\}$. This approximation has been shown to perform well in a number of ABC implementations for which exact likelihood computations are possible (e.g., Jabot & Chave 2009).

A number of improvements of this basic ABC scheme have been developed in recent years (Marin et al. 2012) and implemented in various softwares (Wegmann et al. 2010, Csilléry et al. 2012, Jabot et al. 2013), while ABC implementations have progressively diffused in various biological fields and beyond (Wilkinson 2009, Beaumont 2010). A major remaining challenge for ABC is its computational burden. Many ways to speed up the ABC procedure are available, including the sequential improvements of the approximate posterior distribution (ABC-SMC, Sisson et al. 2007, Beaumont et al. 2009, Toni et al. 2009, Lenormand et al. 2013, Albert et al. 2014), the coupling to Markov chain Monte Carlo (ABC-MCMC, Marjoram et al. 2003, Wegmann et al. 2009) or the use of prior information to perform an ABC-based re-calibration (Lagarrigues et al. 2014), but even these improved schemes typically require tens of thousands to millions of model runs, which is generally unreachable for models whose simulations take more than several minutes of computing time.

To further speed up ABC implementations for computationally-demanding models, it has been recently proposed to use simpler statistical models instead, called emulator or meta-model, that will approximate the outputs of the complex model using various interpolation techniques (Henderson et al. 2009, Jandarov et al. 2013, Meeds and Welling 2014, Wilkinson 2014). Concretely, the coupling of emulation techniques with ABC, hereafter called ABC-Emulation, consists in 1) simulating a small number of model realizations with various parameter values, 2) fitting the emulator on these model realizations, and 3) performing ABC using the emulator instead of the original model.

Meta-modelling has a long history in computer science (Sacks et al. 1989, Santner et al. 2003) and a number of emulation techniques can be mobilized depending on the specificities of the models to be emulated (O'Hagan 2006, Levy and Steinberg 2010). Meta-modelling has been used in particular to perform Bayesian calibration (O'Hagan 2006) and optimization (Jones 2001, Picheny and Ginsbourger 2014). More recent contributions have demonstrated how some emulation techniques could be mobilized in an ABC framework (Henderson et al. 2009, Jandarov et al. 2013, Meeds and Welling 2014, Wilkinson 2014). They have however not provided a detailed comparison of the pros and cons of various emulation strategies nor of the gain brought by ABC-Emulation approaches compared to classical ABC ones. In this context, the aim of this contribution is 1) to assess the gain in computing cost provided by ABC-Emulation, 2) to compare various emulation strategies, and 3) to provide guidelines for potential users. We will compare the efficiency of two well-known emulation techniques – local regressions (Cleveland 1979) and Gaussian processes (Matheron 1963, Rasmussen and Williams 2006), in the context of ABC applications, that is in the presence of a trend between model parameters and summary statistics and of substantial stochasticity in model outputs. We will further explore the influence of the number of initial model realizations on the quality of the emulation-based inference. We will compare emulation strategies based either on the emulation of the summary statistics of model outputs, or on the principal component axes (PCA) of these summary statistics. We will finally explore the use of emulation techniques within a sequential ABC framework (Lenormand et al. 2013). The ABC-Emulation techniques detailed in this contribution have been included in the R package "EasyABC" (Jabot et al. 2013).

## 2. Material and Methods

### 2.1. The strategy of ABC-Emulation

ABC-Emulation consists in four steps. First, a small number *n* of design points are drawn in the parameter space of the model using any sampling scheme. We will here make use of a latin hypercube sampling scheme to cover the parameter space efficiently (Carnell 2012), but alternative schemes could also be used. Importantly, these design points do not need to be drawn in the prior distribution of the model parameters. They only need to provide a good coverage of this distribution so that the approximation of the model works well in all areas of the parameter space. Second, the design points are used to perform *n* realizations of the original model and the *n* model outputs are recorded, where the term "output" represents the set of summary statistics that are computed on the simulation result. Third, an emulator is fitted to these *n* outputs. Fourth, a standard ABC procedure is launched using the emulator instead of the original model. Importantly, this fourth step can also make use of any methodological improvements already developped in the litterature on ABC, such as sequential schemes or post-processing treatments (Marin et al. 2012). A detailed algorithmic presentation of ABC-Emulation is presented in Table 1.

**2.2. Emulation based on local regressions**

The first type of emulator tested consists in performing locally weighted regressions (Cleveland 1979). For a point *x* in the space of model parameters values, this locally weighted regression consists in 1) selecting the *αn* closest design points, where *n* is the number of design points for which the set of summary statistics *v* to be emulated has been computed using the original model and *α* is the proportion of design points that are used – this proportion was chosen with a cross-validation procedure by minimizing the root mean square error of the local regression predictions. This lead to *αn* values that were generally close to 50; 2) computing the weights of the selected design points – here we used a tricubic weight function following Cleveland (1979), that is equal to $(1-(d/d\_max)^3)^3$, where *d* is the distance between the design point and *x* and *d*_max is the maximal distance between the selected design points and *x*; 3) performing a multiple linear weighted regression to predict the value of each composant $v_i(x)$ of *v* at point *x*, using as explanatory variables the model parameters – here we used a polynomial weighted regression of degree 2 following Cleveland (1979): $v_i(\boldsymbol{x})=\beta_{i0} + \Sigma_j\, \beta_{ij}\, x_j + \Sigma_{j<k}\, \beta_{ijk}\, x_j x_k$; 4) recording the remaining non-explained stochasticity at point $\boldsymbol{x}$ by computing the weighted covariance matrix of the residuals of the weighted regression. Emulation-based model realizations at point $\boldsymbol{x}$ are then multivariate normal simulations with mean $\boldsymbol{v}(\boldsymbol{x})$ and covariance matrix equal to the empirical covariance matrix of the residuals of the local regression. We compared inference results based on the emulation of untransformed summary statistics, and on the emulation of the principal component axes of the summary statistics. Seven values of $n$ were compared: 100, 200, 500, 1000, 2000, 5000 and 10000. We further assessed the performance of deterministic emulators in which emulation-based model realizations at point $\boldsymbol{x}$ were equal to $\boldsymbol{v}(\boldsymbol{x})$.

**2.3. Emulation based on Gaussian processes**

The second type of emulator tested was based on Gaussian processes (Rasmussen and Williams 2006, Diggle and Ribeiro 2007). A Gaussian process is a random field $Z(\boldsymbol{x})$ with zero mean and covariance $\sigma^2\, R(\boldsymbol{x},\boldsymbol{y})$ where $R(\boldsymbol{x},\boldsymbol{y})$ is a kernel function to be specified. We here compared five classical kernel functions: Gaussian, exponential, matérn 3/2, matérn 5/2 and power-exponential. Since we were interested in stochastic models, we further used a nugget term (Gramacy and Lee 2012) which is an additional noise term at each design points, following a centred normal distribution with variance $\tau^2$. Gaussian processes were fitted using the R package "DiceKriging" (Roustant et al. 2012) for each summary statistic independently. We also used a variant of this approach in which the principal component axes of the summary statistics were fitted.

Gaussian processes are the most widely used emulators in the meta-modelling literature, although they become computationally-demanding to fit when the number of design points becomes larger than a few hundreds. In the examples below, we were able to fit Gaussian processes up to $n$=500 design points and got saturated memory problems for $n$ equal to 1000 and above. Therefore, only three values of *n* were compared: 100, 200 and 500.

**2.4. Measuring the inference quality of emulation-based strategies**

We performed the analyses on a model of ecological community dynamics (Jabot 2010) that makes use of four parameters and outputs four summary statistics. This model is hereafter referred to as the "community model". It is implemented in the R package "EasyABC" (Jabot et al. 2013) and is thus convenient to manipulate and available for the research community. This model is sufficiently quick to simulate to perform standard ABC procedures, so that it enables to compare the accuracy of emulation-based and standard ABC. We first performed 100,001 realizations of the community model using uniform prior distributions on each model parameter. Prior distributions for parameters $\ln(I)$, $h$, $\ln(A)$, and $\ln(\sigma)$ were [3;5], [-25;125], [ln(0.1);ln(5)] and [ln(0.5);ln(25)] respectively, following Jabot (2010). We randomly selected 100 of these realizations to serve as 100 virtual datasets, and performed, for each of these 100 virtual datasets, a standard ABC inference based on the 100,000 remaining simulations. The 100 resulting approximate posterior distributions serve as reference posterior distributions to measure the inference quality of emulation-based strategies. The inference quality of an emulation-based strategy was then measured as the similarity between the emulation-based and the reference posterior distributions. This posterior similarity was computed for each virtual dataset as the $L_2$ distance between the reference and the emulation-based posterior distributions. To compute this $L_2$ distance, we divided each support of the parameter prior distribution into 6 equally-sized bins, leading to a grid of $6^4=1296$ cells, and we computed in each cell the squared differences between the weights of the reference and the emulation-based posterior distributions (Lenormand et al. 2013). We report in the result section and in the figures the median of the $L_2$ distance among the 100 virtual datasets to summarize the inference quality of a particular emulation strategy. To serve as a comparison, we further launched another set of 100,000 model realizations to assess the speed of convergence of the posterior distribution based on a standard

ABC procedure. Emulation-based and standard ABC procedures were compared with and without using the post-processing treatment based on local regressions proposed by Beaumont et al. (2002).

**2.5 Use of an emulator in a sequential ABC framework**

We finally explored whether emulation techniques could enable to further speed-up sequential ABC techniques. Sequential ABC consists in performing multiple steps of exploration of the parameter space. Starting with a random sampling exploration of the prior distribution, as in standard ABC, sequential ABC procedures then make use of previous ABC simulations to perform an importance sampling of the parameter space. This enables to sample more intensively areas of the parameter space with large posterior densities (Doucet et al. 2001, Marin et al. 2012).

Concretely, our emulation-based sequential ABC algorithm consisted in 1) performing $n$ initial model realizations at $n$ design points, randomly drawn in the prior distribution, using a latin hypercube sampling design (Carnell 2012); 2) fitting an emulator on these $n$ model realizations – we used an emulator based on local regressions and $n$=100; 3) performing a sequential ABC procedure using the emulator, and starting from the $n$ design points – we used the sequential algorithm proposed by Lenormand et al. (2013) that was shown to outperform previously proposed algorithms with default tuning parameters $\alpha$=0.5 and $p_{accmin}$=0.05 ; 4) performing an importance sampling of $n$ new points in the parameter space based on the results of the sequential ABC procedure using the emulator, and simulating the original model on these $n$ new points – we used the importance sampling scheme of Lenormand et al. (2013); 5) repeating steps 2 to 4 a predefined number of times $K$. In these subsequent steps $i$, all the previously simulated model realizations ( $n * i$ ) are used to fit the emulator, but only the $n$ design points which lead to the closest to data summary statistics are used to initialize the sequential ABC procedure based on the emulator. A detailed algorithmic presentation of the ABC-Emulation sequential algorithm is presented in Table

2.

To assess the gain in computing speed provided by the use of an emulator, we computed at the end of each step the maximal distance *d* between the target summary statistics and the summary statistics simulated with the original model, of the *n*=100 closest simulations. We compared this maximal distance to the one obtained with the sequential algorithm of Lenormand et al. (2013).

All analyses were performed with the R software (R Core Team 2014). And routines to perform the ABC-Emulation techniques tested in this contribution have been included in the R package "EasyABC" (Jabot et al. 2013).

## 3. Results

### 3.1. Comparing the different emulation strategies

Our first finding was that stochastic emulators performed largely better than deterministic ones. When the ABC rejection scheme was used, the median posterior similarity among the 100 virtual datasets was almost twice smaller for stochastic emulators compared to the deterministic ones, when a low number (below 500) of model simulations were used to fit the emulator (Fig. 1A). The performance of stochastic and deterministic emulators was similar for larger numbers of model simulations. When the post-processing treatment of Beaumont et al. (2002) was used (Fig. 1B), the superiority of stochastic emulators was even larger, with a median posterior similarity two to three times smaller for stochastic emulators compared to deterministic ones (Fig. 1B). Reproducing the stochasticity of the original model is thus key to get a good approximation of the posterior distribution.

Our second finding was that the gain brought by emulation (Fig. 1A) was largely reduced when the post-processing treatment of Beaumont et al. (2002) was used in concert (Fig. 1B). This is not surprising since the post-processing treatment proposed by Beaumont et al. (2002) already uses information on the local relationship between parameter values and summary statistics. The emulation-based strategy based on local regressions still enabled to get the same inference quality (measured as the median posterior similarity) for about twice less model simulations, when the number of model simulations was between 500 and 2000 (Fig. 1B). The gain was then negligible or null for larger number of simulations. This means that this emulation-based strategy is likely to be useful only for very computationally costly models for which no more than a few hundreds of simulations are manageable.

Our third finding was that the type of emulator used was not affecting much the results (Fig. 2). Emulations based on Gaussian processes were not found to clearly outperform emulations based on local regressions, whatever the type of covariance kernel used. Since emulations based on Gaussian processes become computationally problematic when the number of design points is larger than a few hundreds, we recommend the use of local regressions which do not have such problems and perform similarly or even better (Fig. 2). Finally, the emulation of the principal component axes of the summary statistics instead of the summary statistics themselves was not found to improve inference quality (Fig. 2).

**3.2. Performance of an emulation-based sequential ABC algorithm**

The use of emulators based on local regressions in the sequential ABC framework enabled to speed up the convergence of simulations towards the target summary statistics by roughly a factor of two (Fig. 3). Emulation-based sequential ABC is therefore likely to be an interesting technique for a much wider array of applications than emulation-based standard ABC which was shown to bring a speed up in computing time only for a small range of simulation numbers (Fig. 1B). This algorithm of emulation-based sequential ABC was included in the R package "EasyABC" (Jabot et al. 2013), as well as the other emulation-based standard ABC algorithm.

## 4. Discussion

Meta-modelling has already been used in ecology to upscale spatially and temporally costly simulators (Urban 2005, Marie and Simioni 2014). It has also been used to perform sentivity analysis or pointwise inference of such simulators (Kwon and Hudson 2010, Coutts and Yokomizo 2014). Its use in the context of approximate Bayesian computing (ABC) has been pioneered by a handful of studies which have demonstrated its potential (Henderson et al. 2009, Jandarov et al. 2013, Meeds and Welling 2014, Wilkinson 2014). There is however presently little guidelines on how to choose among the variety of emulation strategies (Levy and Steinberg 2010). This study aimed at comparing various emulation strategies in the context of ABC. Based on the studied community model, we first evidenced that the use of stochastic emulation strategies was critical to get good inference results (Fig. 1). We then demonstrated that emulator types had little impact on the efficiency of the ABC-Emulation inference (Fig. 2). We therefore recommend the use of local regressions instead of Gaussian processes which have been used in most emulation-based ABC studies so far (Meeds and Welling 2014, Wilkinson 2014). Local regressions present the double advantage of enabling to model the covariances among multiple summary statistics and of not posing any computational problem when the number of design points to fit the meta-model becomes larger than a few hundreds.

Another important finding of our study was that most of the gain brought by the use of emulators disappeared when used in combination with the post-processing local regression adjustment proposed by Beaumont et al. (2002) (Fig. 1B). This is not so surprising since this post-processing treatment is actually very close in spirit to a meta-modelling approach.

In contrast, we showed that emulation could be beneficial in the context of sequential ABC schemes, enabling to gain about a factor of 2 in computing cost (Fig. 3). Meeds and Welling (2014) and Wilkinson (2014) already proposed to use sequential schemes. There is however a major difference between their and our schemes. In these previous studies, the quality of the emulator was sequentially improved, and this emulator was then used to compute the posterior distribution of the parameters. In contrast, in our sequential scheme, the emulator is only used to design efficient importance samplings which are sequentially improved, but it is not used to replace the original model in the computation of the posterior distribution. This leads to another advantage of our emulation-based sequential ABC scheme: the discrepancy between the meta-model and the original model can only result in sub-optimal importance samplings which will slow down the convergence of the posterior distribution, but will not bias it.

Our findings are based on a specific simulation model of ecological community dynamics (Jabot 2010). The benefits of using this particular model was that it is sufficiently quick to simulate so that standard ABC inferences can be performed for the sake of comparison. This model is further available in the R package "EasyABC" so that the analyses can be transparently reproduced. Still, the generality of these findings should be further assessed on other case studies. Both the emulation-based standard and sequential ABC inference routines examined in this study have been implemented in the R package "EasyABC" so as to ease the assessment of ABC-Emulation performance in other case studies.

## Acknowledgements

FJ and ND were supported by the RNSC network "MEXICO" and the Irstea INDECO project Dynindic. GL was funded by the French Environment and Energy Management Agency (ADEME)

## References

-Albert, C., Künsch, H.R., Scheidegger, A. (2014). A simulated annealing approach to approximate Bayes computations. *Statistics and Computing*.

-Beaumont, M. A. (2010). Approximate Bayesian computation in evolution and ecology. *Annual review of ecology, evolution, and systematics*, *41*, 379-406.

-Beaumont, M. A., Zhang, W., & Balding, D. J. (2002). Approximate Bayesian computation in population genetics. *Genetics*, *162*(4), 2025-2035.

-Beaumont, M. A., Cornuet, J. M., Marin, J. M., & Robert, C. P. (2009). Adaptive approximate Bayesian computation. *Biometrika*, asp052.

-Carnell, R. (2012). lhs: Latin hypercube samples. *R package version 0.1*.

-Cleveland, W. S. (1979). Robust locally weighted regression and smoothing scatterplots. *Journal of the American statistical association*, *74*(368), 829-836.

-Coutts, S. R., & Yokomizo, H. (2014). Meta-models as a straightforward approach to the sensitivity analysis of complex models. *Population Ecology*, *56*(1), 7-19.

-Csilléry, K., François, O., & Blum, M. G. (2012). abc: an R package for approximate Bayesian computation (ABC). *Methods in ecology and evolution*, *3*(3), 475-479.

-Diggle, P., & Ribeiro, P. J. (2007). *Model-based geostatistics*. Springer.

-Doucet, A., De Freitas, N., & Gordon, N. (2001). An introduction to sequential Monte Carlo methods. In *Sequential Monte Carlo methods in practice* (pp. 3-14). Springer New York.

-Gramacy, R. B., & Lee, H. K. (2012). Cases for the nugget in modeling computer experiments. *Statistics and Computing*, *22*(3), 713-722.

-Grimm, V., & Railsback, S. F. (2005). Individual-based Modeling and Ecology:(Princeton Series in Theoretical and Computational Biology).

-Grimm, V., Revilla, E., Berger, U., Jeltsch, F., Mooij, W. M., Railsback, S. F., ... & DeAngelis, D. L. (2005). Pattern-oriented modeling of agent-based complex systems: lessons from ecology. *science*, *310*(5750), 987-991.

-Henderson, D. A., Boys, R. J., Krishnan, K. J., Lawless, C., & Wilkinson, D. J. (2009). Bayesian emulation and calibration of a stochastic computer model of mitochondrial DNA deletions in substantia nigra neurons. *Journal of the American Statistical Association*, *104*(485).

-Jabot, F. (2010). A stochastic dispersal-limited trait-based model of community dynamics. *Journal of theoretical biology*, *262*(4), 650-661.

-Jabot, F., & Chave, J. (2009). Inferring the parameters of the neutral theory of biodiversity using phylogenetic information and implications for tropical forests. *Ecology letters*, *12*(3), 239-248.

-Jabot, F., Faure, T., & Dumoulin, N. (2013). EasyABC: performing efficient approximate Bayesian computation sampling schemes using R. *Methods in Ecology and Evolution*, *4*(7), 684-687.

-Jandarov, R., Haran, M., Bjørnstad, O., & Grenfell, B. (2014). Emulating a gravity model to infer the spatiotemporal dynamics of an infectious disease. *Journal of the Royal Statistical Society: Series C (Applied Statistics)*, *63*(3), 423-444.

-Jones, D. R. (2001). A taxonomy of global optimization methods based on response surfaces. *Journal of global optimization*, *21*(4), 345-383.

-Kwon, H. Y., & Hudson, R. J. (2010). Quantifying management-driven changes in organic matter turnover in an agricultural soil: An inverse modeling approach using historical data and a surrogate CENTURY-type model. *Soil Biology and Biochemistry*, *42*(12), 2241-2253.

-Lagarrigues, G., Jabot, F., Lafond, V., & Courbaud, B. (2014). Approximate Bayesian computation to recalibrate individual-based models with population data: Illustration with a forest simulation model. *Ecological Modelling*.

-Lenormand, M., Jabot, F., & Deffuant, G. (2013). Adaptive approximate Bayesian computation for complex models. *Computational Statistics*, *28*(6), 2777-2796.

- Levy, S., & Steinberg, D. M. (2010). Computer experiments: A review. *AStA Advances in Statistical Analysis*, *94*(4), 311-324.

- Marie, G., & Simioni, G. (2014). Extending the use of ecological models without sacrificing details: a generic and parsimonious meta-modelling approach. *Methods in Ecology and Evolution*, *5*(9), 934-943.

- Marin, J. M., Pudlo, P., Robert, C. P., & Ryder, R. J. (2012). Approximate Bayesian computational methods. *Statistics and Computing*, *22*(6), 1167-1180.

- Marjoram, P., Molitor, J., Plagnol, V., & Tavaré, S. (2003). Markov chain Monte Carlo without likelihoods. *Proceedings of the National Academy of Sciences*, *100*(26), 15324-15328.

- Matheron, G. (1963). Principles of geostatistics. *Economic geology*, *58*(8), 1246-1266.

- Meeds, E., & Welling, M. (2014). GPS-ABC: Gaussian process surrogate approximate Bayesian computation. *arXiv preprint arXiv:1401.2838*.

- O'Hagan, A. (2006). Bayesian analysis of computer code outputs: a tutorial. *Reliability Engineering & System Safety*, *91*(10), 1290-1300.

- Picheny, V., & Ginsbourger, D. (2014). Noisy kriging-based optimization methods: A unified implementation within the DiceOptim package. *Computational Statistics & Data Analysis*, *71*, 1035-1053.

- Rasmussen, C. E. (2006). Gaussian processes for machine learning.

- R Core Team (2014). R: A language and environment for statistical computing. R Foundation for Statistical Computing, Vienna, Austria. URL http://www.R-project.org/.

- Roustant, O., Ginsbourger, D., Deville, Y. (2012). DiceKriging, DiceOptim: two R packages for the analysis of computer experiments by kriging-based metamodelling and optimization. *Journal of Statistical Software, 51*(1),1-55.

- Sacks, J., Welch, W. J., Mitchell, T. J., & Wynn, H. P. (1989). Design and analysis of computer experiments. *Statistical science*, 409-423.

-Santner, T. J., Williams, B. J., & Notz, W. (2003). *The design and analysis of computer experiments*. Springer.

-Sisson, S. A., Fan, Y., & Tanaka, M. M. (2007). Sequential monte carlo without likelihoods. *Proceedings of the National Academy of Sciences*, *104*(6), 1760-1765.

-Sousa, V. C., Fritz, M., Beaumont, M. A., & Chikhi, L. (2009). Approximate Bayesian computation without summary statistics: the case of admixture. *Genetics*, *181*(4), 1507-1519.

-Toni, T., Welch, D., Strelkowa, N., Ipsen, A., & Stumpf, M. P. (2009). Approximate Bayesian computation scheme for parameter inference and model selection in dynamical systems. *Journal of the Royal Society Interface*, *6*(31), 187-202.

-Urban, D. L. (2005). Modeling ecological processes across scales. *Ecology*, *86*(8), 1996-2006.

-Wegmann, D., Leuenberger, C., & Excoffier, L. (2009). Efficient approximate Bayesian computation coupled with Markov chain Monte Carlo without likelihood. *Genetics*, *182*(4), 1207-1218.

-Wegmann, D., Leuenberger, C., Neuenschwander, S., & Excoffier, L. (2010). ABCtoolbox: a versatile toolkit for approximate Bayesian computations. *BMC bioinformatics*, *11*(1), 116.

-Wilkinson, D. J. (2009). Stochastic modelling for quantitative description of heterogeneous biological systems. *Nature Reviews Genetics*, *10*(2), 122-133.

-Wilkinson, R. D. (2014). Accelerating ABC methods using Gaussian processes. *arXiv preprint arXiv:1401.1436*.

**Table 1. ABC-Emulation algorithm**

1. Sample $n$ design points $\theta_i$ in the support of the prior distribution $\pi(\theta)$

2. Simulate the original model at the $n$ design points $x_i \sim f(x_i|\theta_i)$ and compute the summary statistics of the simulations $S(x_i)$

3. Fit a stochastic emulator g on the $n$ sets of simulated summary statistics $S(x_i) = g(\theta_i)$

4. Use the emulator g instead of the original model in a standard ABC scheme*, that is:

    a) Sample $N$ points $\theta_j$ in the prior distribution $\pi(\theta)$ (with $N \gg n$)

    b) Simulate the emulated summary statistics $S(x_j) \sim g(S(x_j)|\theta_j)$

    c) Compute the distance between the data and the simulations $\rho(S(x_j),S(y))$

    d) Keep the closest to data simulations such that $\rho(S(x_j),S(y)) < \varepsilon$ to form the approximate posterior distribution.

*Note that the rejection-based ABC algorithm has been used here for simplicity. Any other ABC scheme can also be used in step 5 (see Jabot et al. 2013 for a description of ABC schemes currently implemented in the R package "EasyABC").

**Table 2. ABC-Emulation sequential algorithm**

1) Sample $n$ design points $\theta_i$ in the prior distribution $\pi(\theta)$

2) Simulate the original model at the $n$ design points $x_i \sim f(x_i|\theta_i)$

3) Compute the summary statistics of the simulations $S(x_i)$ and the distance between the simulations and the data $\rho_i = \rho(S(x_i),S(y))$

4) Set the initial particle weights $w_i=1$

5) Build a reference table $T=\{(\theta_i,S(x_i),\rho_i,w_i)\}$ of previous simulations of the original model that will be used to fit the emulator

6) For step $k = 1$ to $K$:

   a) Fit a stochastic emulator $g_k$ using all the particles of the reference table $T$: $S(x_i) = g_k(\theta_i)$

   b) Keep the set $S_k$ of the $n$ particles $\theta_j$ of the reference table $T$ with the smallest $\rho_j$

   c) Perform the sequential ABC algorithm of Lenormand et al. (2013) using the set $S_k$ of weighted particles as the starting set and the emulator $g_k$ instead of the original model f. This will produce a final set of $n$ weighted particles $FS_k=\{(\theta_l,g_k(\theta_l),w_l)\}$

   d) Use this final set $FS_k$ to perform an importance sampling $IS_k$* of $n$ new weighted particles $(\theta_m,w_m)$

   e) Simulate the original model for these $n$ new particles $x_m \sim f(x_m|\theta_m)$ and compute their distance to the data $\rho_m = \rho(S(x_m),S(y))$

   f) Add these new weighted particles to the reference table $T$

7) Among the particles of the reference table $T$, keep the closest to data particles such that $\rho(S(x_j),S(y)) < \varepsilon$ to form the approximate posterior distribution.

*$IS_k$ is defined as follows:

1) pick a particle $\theta_i$ of the set $FS_k$ with probability $w_i / \Sigma_j w_j$

2) sample a new particle from a multivariate normal distribution $\theta_i' \sim N(\theta_i,\Sigma)$ with $\Sigma$ equal to

twice the weighted empirical covariance matrix of the set $FS_k$

3) set for this new particle $\theta_i'$ a weight $w_i' = \pi(\theta_i') / \Sigma_j (w_j / \Sigma_l w_l)\, d(\theta_i', \theta_j, \Sigma)$ where $d(x,\mu,\Sigma)$ is the density at x of a multivariate normal distribution $N(\mu,\Sigma)$.

Note that it is also possible to repeat step6 until the proportion of new accepted particles falls below a predefined threshold (as in Lenormand et al. 2013), instead of for a predefined number of times K.

**Figure 1. Comparison of emulation-based and standard ABC.** Inference quality of emulation-based strategies using local regressions in (A) a rejection-based ABC inference and (B) an ABC inference using the postprocessing treatment of Beaumont et al. (2002). Black lines with crosses represent the inference quality of standard ABC inference. Red lines with triangles represent the inference quality of emulation-based ABC with deterministic emulators. Green lines with circles represent the inference quality of emulation-based ABC with stochastic emulators.

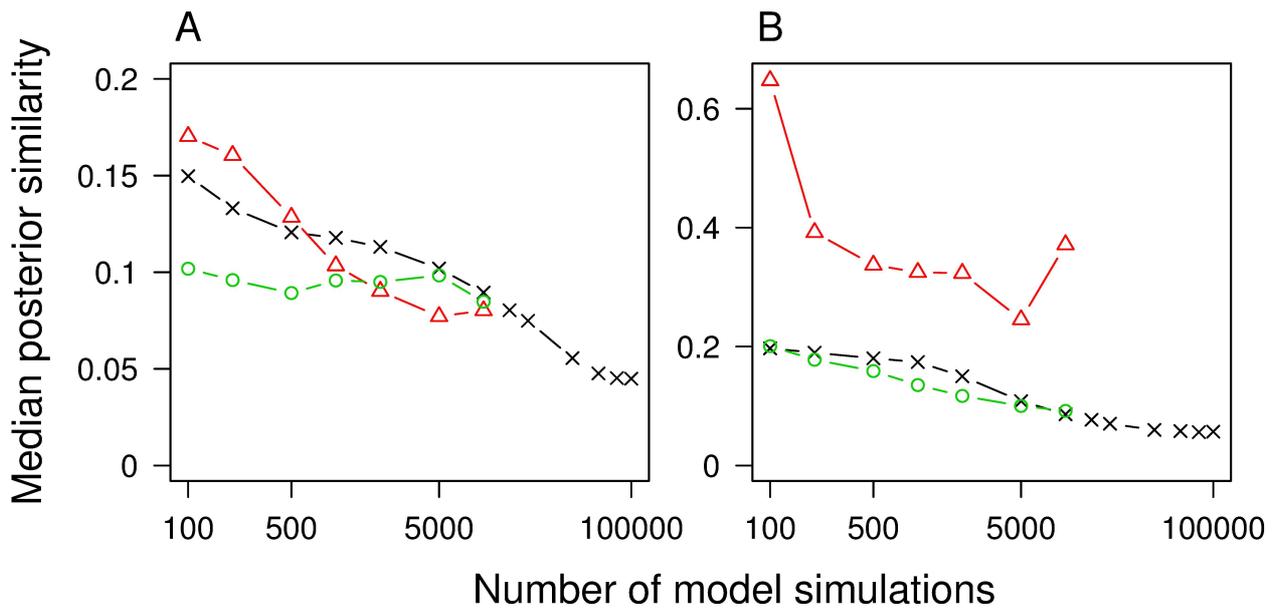

**Figure 2. Comparison of the efficiency of various stochastic emulation strategies.** The green line with circles represent the inference quality of stochastic emulators based on local regressions. Blue crosses represent emulators of principal component axes of the summary statistics based on local regressions. Black symbols represent emulators based on Gaussian processes with Gaussian (squares), exponential (downward triangles), matérn 3/2 (diamonds), matérn 5/2 (upward triangles) and power-exponential (plain squares) covariance kernels. Plain circles represent emulators of principal component axes of the summary statistics based on Gaussian processes with Gaussian covariance kernels. ABC inference was performed using the postprocessing treatment of Beaumont et al. (2002).

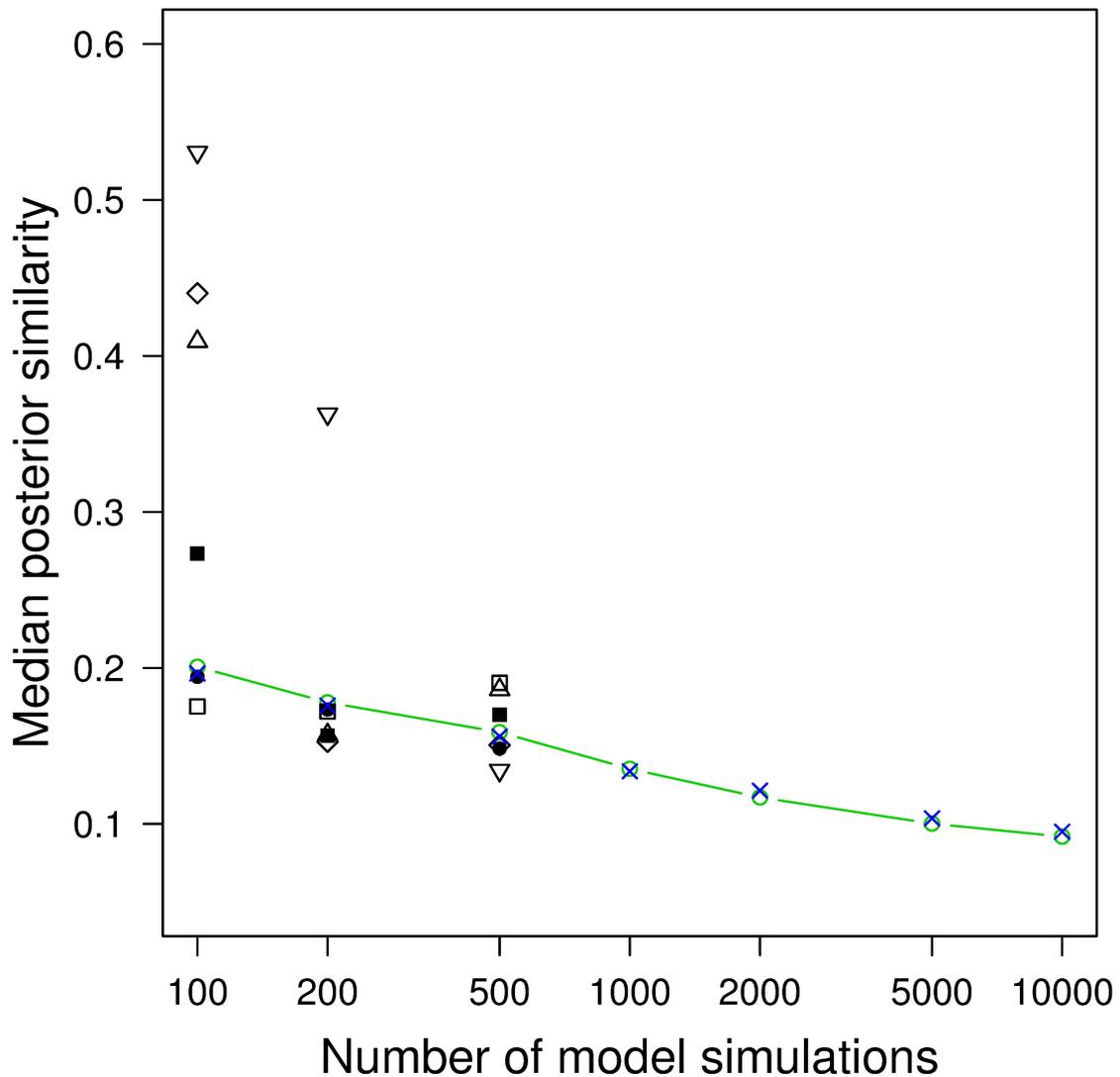

**Figure 3. Efficiency of emulation-based sequential ABC.** Maximal distance *d* between the target summary statistics and the 100 closest simulations for a standard ABC scheme (dotted black line), a sequential ABC scheme using the Lenormand et al. (2013) algorithm (green dashed line) and the emulation-based sequential scheme (blue plain line). The target summary statistics were obtained by simulating the community model with parameters (ln(*I*)=4; ln(*A*)=0, *h*=25; ln($\sigma$)=0.5).

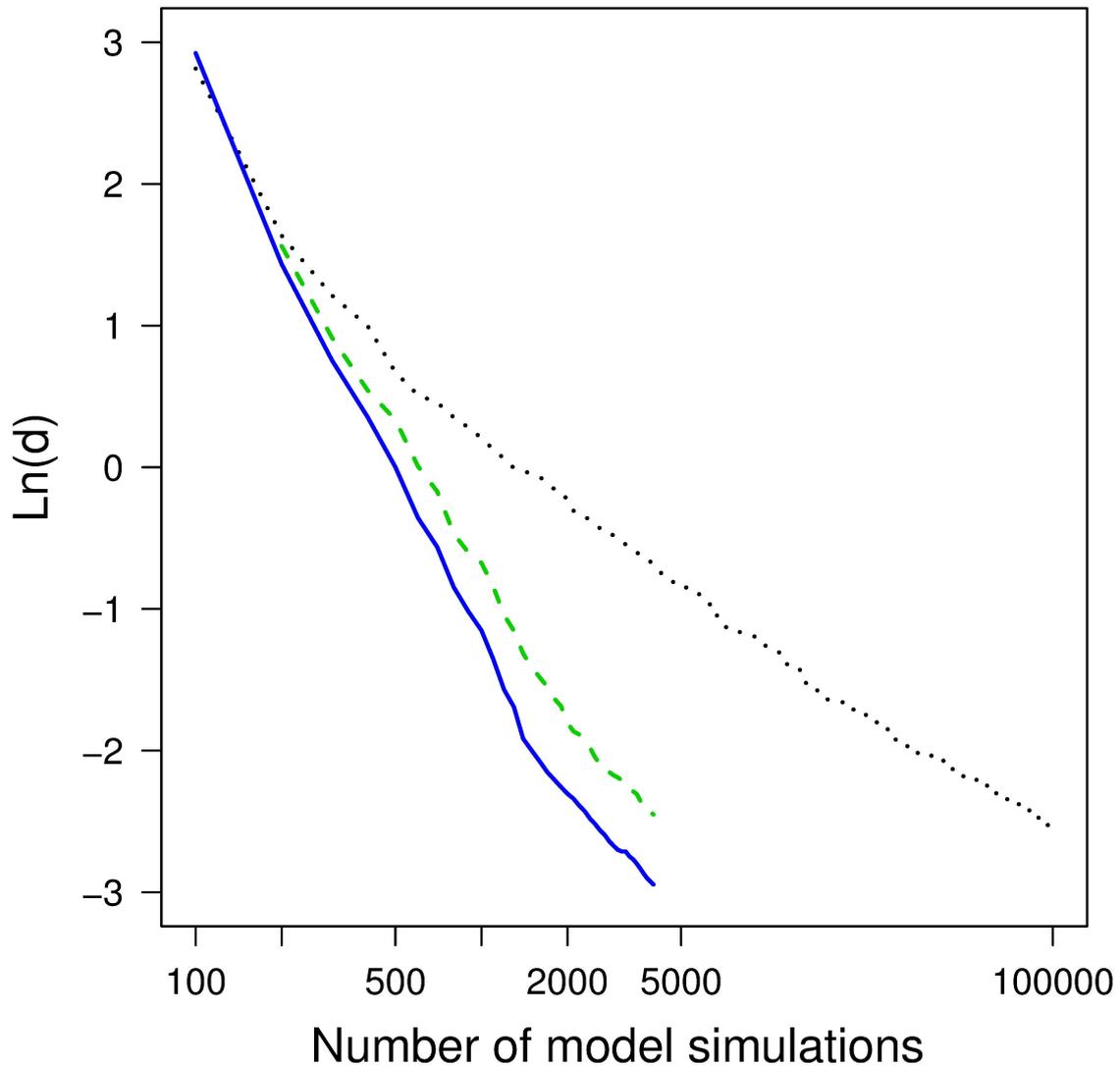